# GPS and the Search for Axions

A. Nicolaidis[1]

Theoretical Physics Department

Aristotle University of Thessaloniki, Greece

**Abstract**: GPS, an excellent tool for geodesy, may serve also particle physics. In the presence of Earth's magnetic field, a GPS photon may be transformed into an axion. The proposed experimental setup involves the transmission of a GPS signal from a satellite to another satellite, both in low orbit around the Earth. To increase the accuracy of the experiment, we evaluate the influence of Earth's gravitational field on the whole quantum phenomenon. There is a significant advantage in our proposal. While the geomagnetic field $B$ is low, the magnetized length $L$ is very large, resulting into a scale $(BL)^2$ orders of magnitude higher than existing or proposed reaches. The transformation of the GPS photons into axion particles will result in a dimming of the photons and even to a "light shining through the Earth" phenomenon.

---

[1] Email: nicolaid@auth.gr

**Introduction**

Quantum Chromodynamics (QCD) describes the strong interactions among quarks and gluons and offers definite predictions at the high energy-perturbative domain. At low energies the non-linear nature of the theory introduces a non-trivial vacuum which violates the CP symmetry. The CP violating term is parameterized by $\theta$ and experimental bounds indicate that $\theta \leq 10^{-10}$. The smallness of $\theta$ is known as the strong CP problem.

An elegant solution has been offered by Peccei – Quinn [1]. A global $U(1)_{PQ}$ symmetry is introduced, the spontaneous breaking of which provides the cancellation of the $\theta$ – term. As a byproduct, we obtain the axion field, the Nambu-Goldstone boson of the broken $U(1)_{PQ}$ symmetry. There are extensive reviews covering the theoretical aspects and the experimental searches for the axion [2, 3, 4, 5, 6, 7].

A general feature of the axion is its two-photon coupling

$$L_{a_{\gamma\gamma}} = -\frac{1}{4} g a F_{\mu\nu}\widetilde{F}^{\mu\nu} = g a \vec{E} \cdot \vec{B} \qquad (1)$$

where $a$ is the axion field, $F_{\mu\nu}\left(\widetilde{F}^{\mu\nu}\right)$ the (dual) electromagnetic field strength tensor and $g$ the photon-axion coupling constant. Accordingly, in the presence of a magnetic field $\vec{B}$, a photon may oscillate into an axion and vice-versa. A prototype experiment in the search for solar axions is CAST experiment, which set the limit for $g < 10^{-10}\,\text{GeV}^{-1}$ [8]. The CAST experiment involves a magnetic field $B=9\text{T}$ and a magnetized region $L=9.3$ m. Therefore, the relevant scale $(BL)^2$ is $(BL)^2 \approx 7000\ \text{T}^2\text{m}^2$. Always in the search for solar axions, a space-based experiment has been proposed, where use is made of the Earth's magnetic field [9]. The weakness of the geomagnetic field $B$ is somehow compensated by the larger $L$ value.

In the present work we suggest using the geomagnetic field in order to study the inverse process, photon-axion transition. We consider a GPS signal travelling from one satellite to another. In the presence of the Earth's magnetic field, the photon may oscillate to an axion and vice-versa. Our proposal involves an increased $(BL)^2$ scale, a higher accuracy and the exploration of a new range of $g$, $m_a$ (coupling constant and axion mass respectively). To further increase the accuracy of our evaluation, we include the effect of Earth's gravitational field on the whole process.

**GPS signal and the influence of Earth's gravitational field**

Global Positioning System (GPS) offers to geodesy position measurements with a millimeter to centimeter-level precision. GPS contributed also to significant advances in geophysics, seismology, atmospheric science and natural hazard science. Besides accurate positioning, all disturbances in the propagation of the transmitted GPS signal from satellite to receiver are mined for information [10]. The GPS system is in effect a realization of Einstein's view of space and time. Indeed, the system cannot function properly without taking into account fundamental relativistic principles [11].

In the presence of the geomagnetic field we may envisage the transition of the GPS signal into an axion. To reach the highest accuracy in the evaluation of the probability P($\gamma \to \alpha$), we must include also the effect of Earth's gravitational field.

The geometry outside a spherical star like the Earth is provided by the Schwarzschild metric

$$d\tau^2 = \left(1 - \frac{R_s}{r}\right)dt^2 - \left(1 - \frac{R_s}{r}\right)^{-1} dr^2 - r^2\left(d\theta^2 + \sin^2\theta \, d\phi^2\right) \quad (2)$$

where $R_s = 2GM$ with $M$ the mass of the star. The relevant scale in our problem, with $R_g$ the radius of the GPS satellite from the Earth's center, is

$$\frac{R_s}{R_g} = \frac{2GM_E}{R_g} \approx 3 \times 10^{-10} \quad (3)$$

Given the smallness of the gravitational strength, we adopt from the very start the metric of a weak gravitational field

$$d\tau^2 = \left(1 - \frac{R_s}{r}\right)dt^2 - \left(1 + \frac{R_s}{r}\right)\left[dx^2 + dy^2 + dz^2\right] \quad (4)$$

The metric is independent of the coordinate t and this implies that the energy is conserved

$$m\left(1 - \frac{R_s}{r}\right)\frac{dt}{d\tau} = p_0 \equiv E = const \quad (5)$$

The relation

$$g^{\mu\nu} p_\mu p_\nu = m^2 \quad (6)$$

provides

$$\frac{E^2}{(1-R_s/r)} - \frac{p^2}{(1+R_s/r)} = m^2 \qquad (7)$$

where we defined $p^2 = \vec{p}^2 = \sum_{i=1}^{3} p_i^2$.

The quantum mechanical phase accumulated by a particle propagated in space-time is given by the invariant quantity [12, 13, 14, 15, 16]

$$d\Phi = p_\mu dx^\mu = E\,dt - \sum_i p_i\,dx^i \qquad (8)$$

Using the relation

$$p_i = m\left(1 + \frac{R_s}{r}\right)\frac{dx^i}{d\tau} \qquad (9)$$

and the relation (5), we obtain

$$p_i = E\frac{(1+R_s/r)}{(1-R_s/r)}\frac{dx^i}{dt} \qquad (10)$$

Subsequently,

$$p = E\frac{(1+R_s/r)}{(1-R_s/r)}\frac{ds}{dt} \qquad (11)$$

where $ds^2 = dx^2 + dy^2 + dz^2$.

The quantum phase acquires the form

$$d\Phi = E\,dt - p\,ds = \left(E\frac{dt}{ds} - p\right)ds = \left[E^2\frac{(1+R_s/r)}{(1-R_s/r)} - p^2\right]\frac{ds}{p} \qquad (12)$$

Equation (7) allows to rewrite

$$d\Phi = m^2\frac{(1+R_s/r)}{p}ds \qquad (13)$$

The above expression is accurate within the weak gravity approach. Notice that in the absence of gravity we obtain

$$d\Phi_0 = \frac{m^2}{p}ds \qquad (14)$$

which is the well-known established result for the Minkowski spacetime.

Working always in the weak gravity limit and ignoring terms $(R_s/r)^2$, we evaluate, using equ. (7)

$$\frac{(1+R_s/r)}{p} \approx \frac{1}{\left[E^2 - \frac{m^2}{(1+R_s/r)}\right]^{1/2}} \approx \frac{1}{\sqrt{E^2-m^2}} - \frac{1}{2}\frac{R_s}{r}\frac{m^2}{(E^2-m^2)^{3/2}}$$

We conclude that

$$d\Phi = \frac{m^2}{\sqrt{E^2-m^2}}\left[1 - \frac{1}{2}\frac{R_s}{r}\frac{m^2}{(E^2-m^2)}\right]ds \qquad (15)$$

In our case we consider a light signal travelling from a satellite at $r=R_g$ to another satellite at $r=R_g$.

The trajectory is almost a straight line and the closest distance to the Earth is denoted by $b$. Then the traveled distance is $s = 2\sqrt{R_g^2 - b^2}$ and

$$\Phi_0 = \frac{m^2}{\sqrt{E^2-m^2}} 2\sqrt{R_g^2 - b^2} \qquad (16)$$

For the second contribution we have to evaluate the integral $\int \frac{ds}{r}$.

Defining $\cos\phi = \frac{b}{r}$ we obtain

$$\int \frac{ds}{r} = \int \frac{d\phi}{\cos\phi} \qquad (17)$$

where $0 \leq \phi \leq \phi_{max}$ with $\cos\phi_{max} = \frac{b}{R_g}$.

We obtain finally

$$\Phi = \frac{m^2}{\sqrt{E^2-m^2}}\left[2\sqrt{R_g^2-b^2} - \frac{m^2}{E^2-m^2}R_s \ln\frac{(1+\omega)}{(1-\omega)}\right] \qquad (18)$$

with

$$\omega = \left[\frac{R_g - b}{R_g + b}\right]^{1/2} \qquad (19)$$

It should be noted that the energy E is the energy measured at infinity $r = \infty$. The energy $E$ and the energy $E_g$ measured at distance $r=R_g$ (the position of the satellites) are connected by

$$E = E_g(1 - R_s/R_g)^{1/2} \qquad (20)$$

The above relation represents the well-known gravitational red shift. Expressing everything in terms of the measured $E_g$ and considering the case $m^2 \ll E^2$ we find the compact expression

$$\Phi = \frac{m^2}{E_g}\left(1 + \frac{R_s}{2R_g}\right)\left(2\sqrt{R_g^2 - b^2}\right) \qquad (21)$$

We conclude that the influence of the Earth's gravitational field can be absorbed into a definition of an effective mass $\mu$

$$\mu^2 = m^2\left(1 + \frac{R_s}{2R_g}\right) \qquad (22)$$

**Photon-axion oscillations**

Consider a GPS signal travelling from a satellite at $r=R_g$ to another satellite at $r=R_g$, both moving at low altitude around the Earth. Let us define as $z$ axis the direction of photon's propagation. The polarization of the photon $\vec{A}$ lies then at the x–y plane. The photon is moving in the presence of the geomagnetic field $\vec{B}$. The component of $\vec{B}$ parallel to the direction of motion does not induce photon-axion mixing. Following eq. (1), the transverse magnetic field $\vec{B}_T$ couples to $A_{\parallel}$, the photon polarization parallel to $\vec{B}_T$ and decouples from $A_{\perp}$, the photon polarization orthogonal to $\vec{B}_T$.

The photon-axion mixing is governed by the following equation:

$$\left(E_g - i\partial_z + \mathbf{M}\right)\begin{pmatrix}A_{\parallel} \\ a\end{pmatrix} = 0 \qquad (23)$$

The 2-dimensional matrix **M** is

$$\mathbf{M} = \begin{pmatrix} -\dfrac{\mu_\gamma^2}{2E_g} & \dfrac{gB_T}{2} \\ \dfrac{gB_T}{2} & -\dfrac{\mu_a^2}{2E_g} \end{pmatrix} \qquad (24)$$

where $\mu^2$ is defined in equ. (22). For a photon, moving in a medium with number density of electrons $n_e$, the photon mass $m_\gamma$ is given by

$$m_\gamma^2 = \frac{4\pi\alpha n_e}{m_e} \qquad (25)$$

The axion mass $m_\alpha$ is not experimentally known. Matrix **M** is diagonalized through the angle $\theta$ with

$$\tan 2\theta = \frac{2 g B_T E_g}{\mu_a^2 - \mu_\gamma^2} \tag{26}$$

Defining

$$D = \frac{1}{2 E_g}\left[(\mu_a^2 - \mu_\gamma^2)^2 + 4 g^2 B_T^2 E_g^2\right]^{1/2} \tag{27}$$

$$\sin 2\theta = \frac{g B_T}{D} \tag{28}$$

we obtain for the probability that a photon converts into an axion after travelling a distance $s$

$$P(\gamma \to \alpha) = \sin^2 2\theta \, \sin^2 \frac{D s}{2} \tag{29}$$

When the oscillatory term in equ. (29) is small, i.e. $\frac{D s}{2} \ll 1$, we obtain the behavior ($L \equiv s$)

$$P_0(\gamma \to \alpha) = \frac{g^2}{4}(B_T L)^2 \tag{30}$$

thus the relevant scale for an experimental setup is $(B_T L)^2$.

Imagine a photon scratching the Earth at $b = R_E$ and becoming an axion at this position. The probability is

$$P(\gamma \to \alpha) = \sin^2 2\theta \, \sin^2\left(\frac{D}{2}\sqrt{R_g^2 - R_E^2}\right) \tag{31}$$

The axion reemerging from the Earth travels to the other satellite. The probability of being detected there like a photon is

$$P(\alpha \to \gamma) = \sin^2 2\theta \, \sin^2\left(\frac{D}{2}\sqrt{R_g^2 - R_E^2}\right) \tag{32}$$

Therefore the probability for light shining through the Earth is given by

$$P = \sin^4 2\theta \, \sin^4\left(\frac{D}{2}\sqrt{R_g^2 - R_E^2}\right) \tag{33}$$

**Conclusions**

We suggest a new experimental technique, where thanks to the Earth's magnetic field, a GPS signal is transformed to an axion particle. There are clear advantages in our proposal.

First, the high accuracy. The distance $L$ is measured with a precision at the millimeter – centimeter level. Second, the large value of the $(BL)^2$ scale. The Earth's magnetic field is of a dipole form with a mean value $B_0 \approx 3 \times 10^{-5}$ T on the Earth's surface. The magnetic field is falling off like $(R_E/r)^3$, where $R_E$ is the radius of the Earth (approximately 6370 km) and $r$ the radial distance from the center of the Earth. To obtain the best available values for the geomagnetic field ($\approx B_0$), the satellites should remain in a low orbit around the Earth. The smallness of the magnetic field, compared, say, to the CAST experiment, is overbalanced by the much larger $L$ value, which may reach $L = 2R_E$. The scale $(BL)^2$ becomes then 140,000 $T^2m^2$, orders of magnitude above existing or proposed values. Third, the energy range of the photons and the possibility to search in an unexplored domain of $g$, $m_\alpha$ values. GPS photons travel with a frequency of approximately 1 GHz. It is appropriate, for our experimental needs, to use a higher frequency of 1 THz, so that we can reach lower values for $m_\alpha$, down to μeV [17] and explore even smaller $g$ values. Finally, our emitter and receiver are in constant motion and therefore the parameter $L$ may vary offering plentiful information. Relying on Einstein's view of space and time, GPS has been established as the ideal tool for geodesy. Next to the relativistic conceptions we included a quantum approach, and we obtained the probability for the transition of a GPS photon to an axion particle in the presence of the Earth's magnetic field. This transition will result in a dimming of the photons and further to light shining through the Earth phenomenon. We may envisage that in the future the long list of scientific disciplines served by GPS will be enriched by particle physics.

**Acknowledgements**

The present work was initiated while I was a visiting scholar at the Center for Axion and Precision Physics, Institute for Basic Science (CAPP – IBS) in Korea. I would like to thank CAPP's director Prof. Yannis Semertzidis for the kind invitation and many enlightening discussions. Aspects of GPS related technology were clarified during a visit at the Laboratoire Astroparticule et Cosmologie (APC) in Paris. I am appreciative of this help provided by APC's director Prof. Stavros Katsanevas. Mr. Dimitris Evangelinos assisted in the typesetting.

## References

[1] R.D. Peccei and H.R. Quinn, *Phys. Rev. Lett.* **38**, 1440 (1977); S. Weinberg, *Phys. Rev. Lett.* **40**, 223 (1978); F. Wilczek, *Phys. Rev. Lett.* **40**, 279 (1978).

[2] G. Raffelt and L. Stodolsky, *Phys. Rev.* **D37**, 1237 (1988).

[3] G. Raffelt, *Nucl. Phys. Proc. Suppl.* 77, 456–461 (1999).

[4] P. Sikivie, in *AIP Conference Proceedings* **805**, 23 (2005).

[5] J. Kim, Talk given at the Corfu Summer Institute 2016, arXiv:1703.03114[hep-ph].

[6] P. Graham, I. Irastorza, S. Lamoreaux, A. Lindner and K. van Bibber, *An. Rev. Nucl. Part. Sci.* **65**, 1(2015), 485–514.

[7] R. Battesti, B. Beltran, H. Davoudiasl, M. Kuster, P. Pugnat, R. Rabadan, A. Ringwald, N. Spooner and K. Zioutas, *Lect. Notes Phys.* **741**, 199–237 (2008).

[8] K. Zioutas *et. al.* (CAST Collaboration), *Phys. Rev. Lett.* **94**, 121301 (2005).

[9] H. Davoudiasl and P. Huber, *Phys. Rev. Lett.* **97**, 141302 (2006).

[10] For a review see Y. Bock and D. Melgar, *Rep. Prog. Phys.* **79**, 106801 (2016).

[11] N. Ashby, Physics Today **55**, pp. 41–72 (2002).

[12] L. Stodolsky, *Gen. Rel. and Grav.* **11**, 391 (1979).

[13] N. Fornengo, C. Giunti, C.W. Kim and J. Song, *Phys. Rev.* **D56**:4, 1895–1902 (1977).

[14] J.G. Pereira and C.M. Zhang, *Gen. Rel. and Grav.* **32**:1633 (2000).

[15] D.V. Ahluwalia and C. Burgard, *Gen. Rel. and Grav.* **28**:1161 (1996); *Phys. Rev.* **D57**, 4724 (1998).

[16] T. Bhattacharya, S. Habib and E. Mottola, *Phys. Rev.* **D59**, 067301 (1999).

[17] S. Borsyanyi *et. al.*, *Nature*, **539**, 69–171 (2016)